# Inventing E-Regulation in the US, EU and East Asia:

# Conflicting Social Visions of the Internet & the Information Society

by

Shalini Venturelli

Associate Professor
International Communication Division
School of International Service
American University
4400 Massachusetts Ave., NW
Washington, DC 20016
Phone: 202-885-1635
Fax: 202-885-2494
Email: sventur@american.edu

Presented at TPRC 2001
29th Research Conference on Information, Communication & Internet Policy
Alexandria, Virginia, October 27-29

Forthcoming in *Telematics & Informatics: An International Journal on Telecommunications & Information Technology*
This paper is drawn from a larger, more extensive study forthcoming from Oxford University Press (2002)

# Inventing E-Regulation in the US, EU and East Asia:

# Conflicting Social Visions of the Internet & the Information Society

by Shalini Venturelli

Associate Professor
International Communication Division
School of International Service
American University
4400 Massachusetts Ave., NW
Washington, DC 20016
Phone: 202-885-1635
Fax: 202-885-2494
Email: sventur@american.edu

This paper attempts to assess the international approach to Internet policy in the context of distinctive socio-political frameworks evolving in the US, the European Union (EU), and East Asia. The comparative review will develop a set of underlying structural models of the Information Society particular to each region, along with an analysis of their defining characteristics in relation to one another. This examination demonstrates how each region, given its regulatory legacy, has elected a different mix of good and bad socio-political choices in public policy for the Internet. Despite the range and diversity of paths to e-regulation suggested in these choices, none adequately addresses the underlying issue of how to promote an innovative society that is open to broad social participation. The paper evaluates principal weaknesses in these regional models of Internet policy and argues the need for re-conceptualizing the cultural, political and economic approach to the new information space of the Internet.

What is needed at this historical juncture in the international system, the paper suggests, is a complete rethinking of the problems and challenges of the new information space in a way that allows its full potential to emerge to the benefit of all sectors of economy and society. The analysis concludes by suggesting the need for a fundamental shift in approach to e-regulation for which two arguments are advanced: first, that the globalization of the information economy and the internationalization of cyberspace have made it imperative that concepts of creativity and innovation be reassessed and repositioned at the center of public policy; and second, that the creative and innovative challenges of the information economy should reform the emphasis on hardware and technological issues by prioritizing the knowledge and creative foundations of society. These arguments suggest the need for a fundamental shift in national and international policy on the appropriate framework for e-regulation in the Information Society.

# Inventing E-Regulation in the US, EU and East Asia:

# Conflicting Social Visions of the Internet & the Information Society

by Shalini Venturelli

This paper attempts to assess the international approach to Internet policy in the context of distinctive socio-political frameworks evolving in the US, the European Union (EU), and East Asia. The comparative review will develop a set of underlying structural models of the Information Society particular to each region, along with an analysis of their defining characteristics in relation to one another. The paper then evaluates principal weaknesses in these regional models and argues the need for re-conceptualizing the cultural, political and economic approach to the new information space of the Internet. The analysis concludes by suggesting the need for a fundamental shift in approach to e-regulation that may offer a more effective design for the democratic and social development of cyberspace.

This paper examines the emerging Information Society in the U.S., Europe, and East Asia, and attempts to show how each region, given a specific regulatory legacy, has elected a different mix of good and bad socio-political choices in public policy. Despite the range and diversity of paths to e-regulation suggested in these choices, it is suggested that none adequately addresses the underlying issue of how to promote an innovative society open to broad social participation. What is needed at this historical juncture in the international system is a complete rethinking of the problems and challenges of the new information space in a way that allows its full potential to emerge to the benefit of all sectors of economy and society. Principal models of the Information Society, including their core characteristics, are identified in three regions of the world, followed by an argument concerning their inadequacies for the social development of cyberspace. This serves as the basis for proposing an alternative e-regulation model with its own distinctive set of international policy implications.

## MODELS OF THE INORMATION SOCIETY

The globalization of the Internet, the digital and wireless revolution, and new media and information systems have provoked the creation of an extensive body of regulation, law, and socio-economic policy in North America, Europe and East Asia. Yet the approaches taken are not identical or even compatible, and reflect profoundly competing conceptions of the information age. Furthermore, each region is characterized by internal conflicts over social and economic priorities and the role of government, reflecting contradictory traditions of regulation, information policy, social models, and political and constitutional constraints. A socio-political examination of some of the principal models of the Information Society competing for influence in public policy in these three regions of the world demonstrates that the form in which information technologies evolve are not intrinsically determined but politically and socially shaped.

### Inventing E-Regulation in the U.S.

The Internet originated in the U.S. from a vision of communication that takes three separate directions, each of which is informed by a particular tradition of regulation and a particular view of society and the state. The special characteristics of the U.S. approach to the Information Society can be explained almost entirely by the dominance of one vision over others in some areas of policy, or from conflicts among the three visions when they overlap or are present in combination in other policies.

The U.S. approach to the information and communications sector and the Information Society emerges from a distinctive social and cultural experience. This experience reflects, at different historical moments, divergent concepts of the individual and fundamental rights, the nature of civil society and the market, and the legitimate scope of state power with respect to each. These social, political, and cultural conceptions can be organized into three traditions of regulation that are tied to separate American theories of the individual, society, and the state (for a more detailed examination of these theories and their links to communication policy and law, see Venturelli, 1998a).

***U.S. Libertarian Model***

The libertarian vision of the Internet and communications technology derives from a concept of society without the state. The foundation of the concept can be traced to the post-feudal, Anglo-Saxon political ideas of John Locke (1690/1960), Thomas Hobbes (1651/1991), and Adam Smith (1776/1986) (for a detailed critique of these ideas and their expression in modern communication regulation, see Venturelli, 1998a). It also informs the notion of communication and exchange of ideas that characterized the underlying working assumption of scientists and engineers who originated the technology of the Internet and wrote its founding codes and protocols (for a brief history of these protocols, see Segal, 1995; and Krol, 1992). In the libertarian view of communication and regulation, the information networks of a society are open and non-proprietary, with strict constraints placed upon state intervention on any grounds, even for, what might be considered, entirely rational aims, such as the protection of minors. It is suggested here that from the standpoint of a minimal state tradition (see Venturelli, 1998a) that libertarian conceptions of information networks represent, the rules of conduct within the system are only regarded as legitimate if they evolve from consensus among users and participants, with conflicts resolved largely through negotiation and dispute resolution. Thus the libertarian view of society and communication is averse to control, regulation or monopolization in information space, and supportive of unobstructed processes of interaction among individuals.

Certainly, the intellectual roots of this tradition can be found in Anglo-Saxon political theory, but the evolution of the idea, it is argued, is firmly entrenched in the American historical experience of the perpetual New Frontier which pervades American cultural understanding of the nature and promise of new technology. New media digital and wireless technologies in a networked environment thus become the New Frontier of a different space that must be carved and structured by those who utilize and exploit it, and not by institutional authorities, proprietary regimes, traditions of law, or the coercion of the state. While the libertarian culture of the Internet is now considered to be more descriptive of its genesis than its current structure, this model of the Information Society has persisted in many forms, including the continuing struggle to create an open architecture and platform for information exchange. The debate over an Open Source Code for music, data, and file sharing, and self-regulatory norms in place of legal codes, in more ways than one, represent the resilience of the libertarian paradigm even while it has been gradually overtaken by other, more powerful American models of regulation (for a typical example of a libertarian view of cyberspace, see Godwin, 1998; for a defense of policies premised in libertarian assumptions, see Lessig, 1999)..

***U.S. Public Interest Model***

The public interest model, too, has roots in the American experience of social and political tensions between the rights of individuals and the boundaries of government (see analysis of this tradition in Venturelli, 1998a; for a regulatory history, see Horwitz, 1989). However, the public interest model resolves this tension by arguing for, rather than against, the application of state power to protect individuals from an even greater social threat thought to reside in unaccountable structures of the marketplace. Ranging from the achievements of antitrust law to universal service and consumer protection policies, the public interest tradition in the U.S. attempts to balance the interests of consumers with those of the industry. The essential point to emphasize in the lesson of this model is that, rather than being founded on a diminished belief in market forces, it is founded on the reverse: a faith that perfect markets can, in fact, be achieved by rational regulation. The public interest model assumes that markets can be made to behave as classical liberal theory says they should (see further discussion of this in Venturelli, 1998a, 1997b). To assure the structures of a functioning market in communications, whereby innumerable producers and consumers of a commodity or service are finely balanced so that no single market participant dominates, the tradition applies a complex body of mechanisms. These include: competition law, access rules, universal service in areas of essential need, educational applications and subsidies, government investment in scientific and technological research, privacy protection and protection of minors, unbundling, interconnection, and interoperability requirements, ‘fair use’ and a limited monopoly principle in intellectual property laws, and safety, standards, tariff, and fair trade practices, among many other public interest regulations (see Venturelli, 1998a).

***U.S. Liberal Market Model***

The regulatory tradition in the U.S. that best conforms with liberal economic principles--as classically delineated by Adam Smith (1690/1960), and later by its intellectual exponents in the modern age, such as Hayek (1944/1994)--emphasizes contractual rights and proprietary freedoms for market participants. The principles require a minimum of state intervention in market relations between producers, consumers and distributors of goods and services. Any emergence of monopolies and market distortions would be insufficient grounds for inviting state intervention, since dominant market entities are regarded as evidence of market success, not market failure (see critical examination of the model in Venturelli, 1998a, 1997b). Inversely, the state should also reject all forms of industrial policy to protect or enhance the position of national industries. In short, the liberal market model neither admits the possibility of, nor accounts for, the phenomenon of ‘dysfunctional markets,’ which the U.S. public-interest model is fundamentally geared to address. Withholding state power from intrusion into market imbalances clearly works to the advantage of structures that are the inevitable outcome of unhindered market processes, such as large-scale entities and vertical and horizontal consolidation (Venturelli, 1998a, 1997b). Also termed ‘supply side’ since they are favorable to market leaders, these policies make no legal distinction between the rights of artificial entities and the rights of individuals, a feature of the U.S. regulatory tradition that is unique among Western industrialized democracies. A series of judicial decisions in the nineteenth century conferred upon artificial entities the same political, property, due process and private rights that the U.S. Constitution till then only extended to individual human citizens (e.g., *Munn v. Illinois* [1876]; *Union Pacific Railroad Company v. United States* [1878]; for U.S. case law extending fundamental rights to legal entities, see Killian & Costello (eds.), 1996). Over time, therefore, the speech rights of individuals to be free from the interposition of state preferences, has become legally identical to the speech rights of artificial entities such as media companies and communications industries. Under a system of rights equivalency between human citizens and artificial entities, the latter may claim the same fundamental freedoms in order to preempt government imposed public interest obligations, such as obligations to open their electronic space for the airing of candidates' views during elections, or providing public information to serve the democratic process.

The U.S. liberal market model of the Information Society accounts for several significant features in the design of U.S. policy, with the underlying common thread of governmental restraint in requiring non-commercial obligations from the communications industry. The economic incentive approach to intellectual property (see Venturelli, 2000; 1998a); for example, is tilted toward contractual extraction of the property right in ideas and expression, which is of central importance to the content industry, rather than towards public access rights to information or towards author rights of paternity and integrity in their creative works. Under this system, public access rights and author rights would be seen as a burdensome and harmful imposition of public interest obligations on the content industry, which may lead to higher market costs in the development and dissemination of content (ibid.). Similarly, the absence of strong privacy protection, or the reluctance to impose competition law and open access rules in a consistent and rigorous manner in order to remedy infrastructure or content monopolies, are typical features of the liberal market model. The approach is inclined, instead, to encourage industry self-regulation in media and communications; few, if any, ownership and market share restrictions; elimination of tariff regulations; and abrogation of content regulations, even for those classes of information and knowledge that would serve public-opinion formation or education (ibid.).

The liberal market model differs from the U.S. libertarian model in substantive ways, principally with respect to prioritizing the market and proprietary issues in public policy. The liberal market model argues for a more powerful legal regime guaranteeing contractual and proprietary rights in the marketplace, whereas the libertarian model prioritizes open and unrestricted exchange of ideas and other forms of transactional and non-transactional relations among individuals, irrespective of market or proprietary imperatives.

***Structural Characteristics of U.S. Policy for the Information Society***

Despite the active and contradictory dynamic of three models competing for dominance in public policy, it is nevertheless still possible to identify a few essential trends in the U.S. framework for the Information Society. These trends reflect a mix of principles, ideas and assumptions drawn from three separate regulatory traditions and social visions. Taken together, the emerging choices can be said to comprise a distinctive socio-political regulatory architecture that defines the U.S. approach to e-regulation (prominent examples of these trends can be found in U.S. Congress, 1998, 1996; U.S.

Government, 1999, 1995, 1993; U.S. Internet Council, 2000, among many other policies and laws). Characteristics include:

- A preference for self-regulatory industry codes
- The privileging of contractual law over public law
- Maintaining a low threshold of privacy rights--other than a general protection against state intrusion which does not resolve the issue of industry and commercial intrusion
- State-led trade policy begins to substitute for industrial policy
- A shift from the ‘fair use’ public access model of intellectual property towards the economics incentives model that favors the content industry
- Removing most constraints on vertical and horizontal consolidation of media, infrastructure, and information industries
- Lifting most public interest, non-commercial obligations from the content industry, and from the cable and telecommunications infrastructure industries
- Reaffirming constitutional constraints on content regulation in new media and cyberspace

(For a more detailed examination of these characteristics, see Venturelli, 2000, 1998a)

## **Inventing E-Regulation in the European Union**

Since the early 1980s, the growing integration of the global economy and the expansion of trade liberalization have challenged the very basis of European policy toward the communications sector. Extensive study (see Venturelli, 1998a, also 2000, and 1997a) has show that within the European Union (EU), telecommunications and audiovisual policies, for instance, have evolved as powerful sites of conflict not only between rival political and social interests, but also among rival paradigms of how the EU ought to approach the design of the Information Society: the liberal market model, the public service model, and the nationalist or culturalist model. Initiatives launched by the European Union have challenged national traditions of communications policy and law. Yet these traditions are proving quite resistant and have transformed, in reverse, the Information Society agenda at the European level.

Two issues should be given serious consideration when examining the EU's path to the information age. First, the European liberal market model and public service model should in no way be confused with the U.S. market model or the public interest model, since underlying assumptions of society, state, the individual, and the nature of policy and law in the U.S. and EU are fundamentally different (Venturelli, 1998a, 1997a). Second, just as liberalization has often been oversimplified as a minimum regulatory approach when, in fact, it is most often a re-regulatory approach (Venturelli, 1998a, 1997a, 1997b), so, too, the interventionist or *dirigiste* state is frequently caricatured as a phenomenon of French history, which obscures the real nature of the European approach to liberalization of the information sector (ibid.). Consequently, the EU debate over the role of the state in the communications sector is far less about a choice between intervention and nonintervention, than it is a contest among three principal forms of intervention and which social model of the Information Society ought to prevail in European democracies (ibid.).

***EU Liberal Market Model***

National regulatory traditions in EU member states have historically followed the public service model or the national-cultural model for organizing the structure of communications. Thus the liberal market approach to regulation in the EU does not originate at the level of member states, but within the mandate to create a single market for goods and services granted to the European Commission by the 1957 Treaty of Rome (Commission of the European Communities, 1993b). Yet it was not till the late 1980s and early 1990s that the Commission chose to apply this mandate to the communications sector (see Venturelli, 1998a, 1997a) through the gradual introduction of privatization, deregulation, and competition in audiovisual and telecommunications services (ibid.). During the past decade, the mandate has expanded even further to appropriate the entire framework for the Information Society (ibid.).

The EU's approach to a market model in the information sector favors, for example, liberalization of telecommunications networks, a shift from author rights to exploiter rights in intellectual property laws, and a move toward self-regulatory codes (ibid.). These are policy positions similar to those found in the U.S. liberal market model. However, a number of core departures in assumptions regarding how the market works, the nature of competition, which social needs take priority, and what remedy options are open to the state cause the EU liberal market model to display sharp dissimilarities with its counterpart in the U.S. Most important among these, in terms of the long-term architecture of e-regulation, is the continued emphasis upon and fortification of, public law as opposed to contractual law. In the U.S., contractual law is ascendant in all market relations and the policies that govern them, with deep roots in Anglo-Saxon Common Law. In the EU, the public law tradition regards contractual law as simply one category of private law governing relations between private parties, and in no way a constraint upon public law and constitutional principles whose roots lie in Roman, Germanic, and Napoleonic legal codes (for a critical examination of these traditions and their implications for communication policy, see Venturelli, 1998a and also 2000, 1997a).

As noted in a recent study (ibid.), the European public law tradition, now integral to the European Union constitutional framework, is conceptually and inherently inclined towards privileging the general interest over private, proprietary, or contractual interests (see examples of the general interest approach in Commission of the European Communities, 1999, 1995a, 1994). Under such a socio-legal system, industries are less able to avert the imposition of general interest obligations than they would be in the U.S. where sweeping statutory principles on market processes are rarely undertaken (ibid.). Furthermore, and closely related to the issue of public law, are substantive differences in the concept of 'market competition' and legitimate market behavior. The European Commission has adopted competition principles for the information sector as steps toward sound liberal economic strategy, but because these are guided by a public law framework, the general interest is necessarily invoked in each case where competition becomes an issue (see Commission of the European Communities, 2000a; for analysis of the EU approach to competition, see Venturelli, 1998a). Whereas the U.S. liberal market model regards competition as an autonomic process of the market when the state is excluded and thus not subject to scrutiny as consolidation and imbalances emerge, the EU approach to market competition in content and infrastructure industries is to allow a broader, more active scope for public intervention in bringing about non-distorted, rational structures (Venturelli, 1998a, 2000).

Finally, the EU liberal market model has opted for a different set of regulatory priorities as compared with the U.S. in its attention to promoting transnational networks in data, communication, and content distribution (examples of this focus may be found in Commission of the European Communities, 2000a, 2000b, 1999, 1995a, 1993). This has been accomplished by subordinating other priorities which the U.S. market model would rank higher, such as: removing burdensome public interest regulations and ownership restrictions, lowering the percentage of state share in infrastructure industries, minimizing consumer privacy protection, simplifying licensing rules, and streamlining and making more transparent the regulatory and rule-making process (see U.S. perspective on this, in U.S. Government, 1999 and United States Internet Council, 2000; for analysis, see Venturelli 2000, 1998a).

Judging from their respective design of an Information Society framework, it is evident that the EU and the U.S. are operating under very different political theories and economic assumptions of the how the information marketplace works and by what standards its proper functioning should be evaluated.

***EU Public Service Model***

The public service model that competes against the EU liberal market model and delimits its scope is more indigenous to European societies with roots deeply intertwined with the modern history of the region (for a detailed examination of this tradition and its significance in communication policy, see Venturelli, 1998a, also 2000). Similar to the differences in market models between the U.S. and EU, the public service model should never be confused with the U.S. public-interest model with which it shares few, if any, historical and socio-political affinities. While the U.S. public-interest tradition stems from the experience of farmers and labor unions struggling against unregulated corporate power during the industrial revolution, eventually leading to the contemporary consumer-oriented theory of government responsibility, the EU public service model of regulation is inextricably bound to the constitutions of European democracies and to the fundamental legitimacy of the state. This is why, it is argued, that the notion of interventionism is no longer, nor has ever been, a useful basis for explaining trends in the EU, or differences between EU and U.S. methods for addressing policy challenges in the digital age (see more discussion of this in Venturelli, 1998a).

Assumptions about economy and society in the public service model stress the constitutional obligations of the legal and legislative system in guaranteeing the institutional arrangements for services essential to citizenship and broad citizen participation. Certainly, access to the public communications system has always fallen under the category of services essential to citizenship and is regulated on constitutional grounds rather than by market or consumer imperatives as it is in the U.S (ibid.). Just as a significant body of ideas in Anglo-Saxon political theory supports the U.S. libertarian and market models, so, too, a strong intellectual foundation adds weight to the public service tradition (ibid.). The foundation includes the constitutional, legal, and participatory political principles outlined by Montesquieu (1748/1989), Rousseau (1755/1973), Hegel (1821/1952) and Kant (1784-97/1991). The public service model for the Information Society requires that the governing principle of a modern free society incorporate not merely the rights of private property, contractual freedom, open competition and functioning information markets, but also the rights of citizens—not 'consumers' as individuals are categorized in U.S. communications policy. Citizens' rights under this EU model would include rights to comprehensive information services and to access at all levels in the public communications networks (for examples, see Commission of the European Communities, 2000b, 1994; French Government, 1999a, 1999b; see analysis, Venturelli, 1998a).

The public service approach to e-regulation describes not only the French constitutional tradition of public policy, but also that of several other states, including, for example, Belgium, Italy and Germany (see background in Lasok & Stone, 1987; analysis in Venturelli, 1998a). More to the point, however, the public service model of regulation and the statutory responsibility it implies is now explicitly provided for in two constitutional provisions of the treaties of the European Union: Article 2 of the Treaty of Rome (Commission of the European Communities, 1993b), and Title 1, Article B of the Maastricht Treaty (Commission of the European Communities, 1993b).

These constitutional provisions at the member state level and the EU level underscore the role of government (both national and transnational) in guaranteeing the general welfare of citizens and their access to services essential to participation in society and economy. In terms of Internet regulation, the public service model would stress policies with: universal service guarantees, high standards of privacy protection, recognition of author rights and human rights in intellectual property, stringent rules for competition, content regulation to ensure production and distribution of information essential to public-opinion formation and to education, standards for quality of information networks and services, public investment in research and innovation, mandated applications for social services in employment, health care, and social welfare (see many components of this framework in Commission of the European Communities, 2000a, 2000b, 1999, 1995a, 1994).

Yet serious questions remain regarding the coherence of the EU Information Society framework, since many of the measures adopted under the public service model stand in clear conflict with those adopted under the EU liberal market model (see detailed discussion of this in Venturelli, 2000 and 1998a). While the form and nature of these inconsistencies are not entirely commensurate with those found in the U.S, the pattern of contradictory, self-canceling features of e-regulation in the EU framework parallel those in the U.S. framework.

***EU National-Cultural Model***

The national-cultural model of the Information Society in the EU is a substantive departure from the public service approach to communications policy in terms of the rights of individuals, the industry, and the primary responsibilities of the state. As examined in a recent study (Venturelli, 1998a), the vision of the national or cultural collective as a basis for regulating the information sector is also substantively at odds with the liberal market model which favors greater freedom for investors and producers from state intervention. While certainly not unique in international terms, the national-cultural regulatory tradition is indigenous to European societies and is characterized principally by an emphasis on content policy over infrastructure policy. Communications industries are regulated to require carriage of content that celebrates, supports, and enriches the “national culture,” whose meaning corresponds to the origins and construction of the nation state system (for further analysis of cultural policy as communication policy, see Venturelli, 2001, 1998a, 1998b). This strategic priority diverges from the central aim in a public service approach which is to ensure universal service conditions. The relationship between the liberal market and nationalist models in the politically sensitive content sector is evident in the tensions between liberal versus nationalist initiatives for audiovisual policy, content regulation, and program production and distribution.

It has been argued (Venturelli, 1998a) that a national-cultural model of the Information Society advocates the social community as an expressive rather than a political or economic unity, one that must be sustained in an uncontaminated and pure form. The expressivist conception of communication policy thus defends the idea of culture as its own ‘form,’ which must be freed from external constraints on its development. Such reasoning is present not only in communication policies of all EU member states, but also in the EU constitution and its extensive corpus of laws and regulations for the media and information industries (examples may be found in Article 128 of the Maastricht Treaty, Commission of the European Communities, 1993b, also 1999, 1995b, 1994, 1989; in addition, European Parliament, 1989). Cultural provisions in the EU treaties reinforce the constitutionality of content regulation to protect culture through myriad mechanisms. These include: state intervention in support of indigenous content production; industrial policy strengthening European information industries in both infrastructure and content; promotion of program production which invigorates national identity; and reformulation of ownership regulation to retain European ownership of cultural industries (ibid.).

Despite these serious efforts, the EU has encountered profound difficulties in reconciling the principles of creating a transnational information services market (the liberal market model) with the aims of promoting cultural identity and growth in European content production (the national-cultural model). One solution would be to apply the liberal market model directly to the cultural sphere by treating the latter, from a policy standpoint, as an economic category. As has been noted (Venturelli, 1998a), this answer was first suggested by former Commission President Jacques Delors (Commission of the European Communities, 1993c) as a way to redefine the cultural issue as an economic strategy. Under this approach, the economic basis of cultural production, rather than culture itself, could become the rationale for state intervention, thereby creating more consistency and lessening tensions in international trade relations, especially with the United States (see discussion of this issue in Venturelli, 1998a).

The cultural model of the Information Society has been re-energized, to some extent, as EU policies continue to convert cultural goals into economic strategies firmly grounded in the treaties and in their mandate for a common marketplace in commodities and services. Regulating economic activities whose forms are cultural is now regarded as consistent with EU constitutional and market principles (see Commission of the European Communities, 1999, 1995b). The multimedia and audiovisual policies of the EU, for example (ibid., unite European concern for cultural protection with social policy, economic policy, and technology policy. The foremost objective of the cultural-national model in e-regulation terms, is to strengthen the European content production industry, ensure that new information services contribute to growth and employment, and look to advanced information services as an effective means to address social and cultural development of European citizens. Multimedia services are thus gradually being assimilated to broader industrial as well as cultural/social goals, a process that is regarded as an important way of fortifying the goals of cultural heritage and cultural cohesion (see analysis in Venturelli, 1998a, 1997a).

The EU is now treating the multimedia content sector as an opportunity to design a new legal framework for extending cultural policy goals to the broadband network. Nevertheless, from a liberal market standpoint, the national-cultural model erects a dense barrier of external, noncommercial burdens for new entrants, thereby distorting the processes of actual competition and circumscribing a highly political and nationalist matrix for the information marketplace. This perception has at times led to deep differences and tensions in bilateral and international relations, especially between the EU and the U.S.

***Structural Characteristics of EU Policy for the Information Society***

The simultaneous existence of three divergent approaches to the Information Society confers a unique character to the European communications market that deviates substantially from policies for the digital age evolving in North America or East Asia. This points to a separate and distinctive European path to e-regulation and the Information Society, with the following core attributes:

- Persistence of the European social model and political tradition of public service regulation as reflected in higher levels of protection for individual citizens in cyberspace

- Strategic importance of national identity and preservation of national culture as a fundamental matter of social solidarity, demonstrated in relatively higher levels of content regulation
- Significance of the European public law tradition that constitutionally privileges the general interest and universal access guarantees in information infrastructure and services, over proprietary rights and contractual freedoms
- Treating the Information Society as a transnational integration agenda
- Continued structural resilience of national infrastructure industries, such as telecommunications, in the face of liberalization and the application of the market model to the information sector
- Greater willingness to implement competition laws and statutory regulations in the information sector than is found under U.S. policy

## Inventing E-Regulation in East Asia

OECD estimates indicate that the Asia-Pacific region, which includes the East Asian economies of Japan, South Korea, Taiwan, Hong Kong, Malaysia, Indonesia, Singapore and China, has experienced rapid growth in new information technology in the past few years (OECD, 2000). With 70 million Internet users in a region representing half the world's population, the growth forecast of 200 million users by 2003 will far exceed Internet usage rates in North America and Europe. The region is also experiencing dramatic shifts in economic reform, especially with respect to the information sector, which suggests that the East Asian form of an Information Society is still unfolding and difficult to characterize. Even so, as in the case of the EU and the U.S., it is possible to identify certain underlying models of society, economy and regulation upon which the Information Society and e-regulation framework is likely to be constructed. An outline of at least two such models will be attempted here: the state-led development model, and the liberalized corporate model. To the extent that neither of these can be identified in Europe and North America in exactly or even approximately the form they are found here, East Asia has the historic opportunity to evolve a separate paradigm and international standard for the information age.

***East Asian Development Model***

The policy tradition common to East Asia in the past forty years is typified by government intervention in industry, labor and credit markets (structural aspects of this are described in Yuhn & Kwon, 2000; Tsao, 1985; Nishimizu & Hulten, 1984; Kim & Roemer, 1979). The comparative economic success of the East Asian development model when measured against the record in South Asia or other developing regions can be attributed, not to intervention since that is a common feature of development policy, but to aggressive industrial policy emphasizing: technology transfer to compensate for domestic innovation scarcity; export-led growth to compensate for scarcity of capital; strategic use of scarce capital in high productivity industries; protection of domestic industries from foreign competition, and coordinated cooperation at all levels between the state and industry and among industries and industrial sectors to improve productive capacity (ample documentation of these strategies since the 1960s can be found in many sources, including OECD 2000; Yuhn & Kwon, 2000; Tsao, 1985; Nishimizu & Hulten, 1984; Kim & Roemer, 1979).

Predictably, under this model, telecommunications, broadcasting, and the media technology sectors have been largely state-owned public monopolies, closely integrated into the coordinated government-industry control structures. There was little, if any, scope for stimulating competition and independent innovation to respond to consumer needs and demands (Venturelli, 1999; OECD, 2000). Beginning in the 1980s and through the mid 1990s, the inflexible industrial model of economic development began to show signs of market distortions and economic recession. While there are many explanations for this decline as accounted for in economic research, it is also argued (see Venturelli, 1999) that semi-industrialized countries became unable to look beyond their model to examine their excessive reliance on imitative technologies developed in countries outside the region. Nor did they seriously question the low levels of investment in technological and service innovation, or the simultaneous emphasis on the ill-considered strategy of dependence upon heavy industrial conglomerates. East Asian economies had benefited in the past from wage and price advantages and from reliance on imported technologies; however, this strategy, as argued here, was completely unsuited to an information revolution--not merely a technological revolution--and an information economy.

The first price paid by the East Asian development model, in terms of the structural needs of an information economy, was the telecommunications sector. Telecommunications industries in East Asia were characterized by distortions of all kinds caused by state intervention, barriers to technological and market development, and insufficient regulation to promote competition and consumer interests (Venturelli, 1999; OECD, 2000).

The second cost of the East Asian development model to an information economy was the democratic deficit. It is argued on the basis of extensive study (see Venturelli, 2001, 1999, 1998a), that in order for an information and knowledge society to emerge and flourish, there are several social and political preconditions that must be met. These preconditions include (for a discussion, see Venturelli, 2001, 1998a): (1) a functioning public sphere (print, electronic, digital, broadband) open to broad participation and deliberative engagement among major social groups; (2) a percentage of the public communications system capacity reserved for non-commercial exploitation in order to strengthen the foundations of civil society and associational development; (3) guarantees of citizens' information rights through freedom of information laws, government transparency, and public service obligations for information providers to serve the public-opinion formation process; and (4) access to knowledge, information and an educational system that cultivates independent judgment instead of rote learning.

While the diffusion of information technology may alleviate the margins of the democratic deficit, there is no necessary relation between information technology and participatory democracy which, as argued, is the only form of democracy from which an Information Society can successfully evolve (Venturelli, 1998a). An industrial model of democracy that manages on minimal participation through voting rights and representative government is no longer adequate to the increased creative and knowledge demands of the information age (see a fuller account of this phenomenon in Venturelli, 2001 and 1999). Thus without the institutional and regulatory structures to ensure participation and access to knowledge and information, as well as broader foundations for civil society, the East Asian economies would be unable to develop and exploit information technologies and services in the long run (ibid.). This, in turn, eventually risks endangering their competitive edge in the global economy.

***East Asian Liberalized Corporate Model***

Following the shift toward economic liberalization and democratization in the 1980s and 1990s, the East Asian economies began to implement macro-economic stabilization plans and shifted their industrial policies (see OECD, 2000; Yuhn & Kwon, 2000). The repeal of selective industry promotion laws and reduction in preferential credit, tax concessions and price control regulations were welcomed by international economic institutions such as the IMF and the OECD. But it did not significantly reduce government intervention in guiding the economy nor abolish state-to-industry or industry-to-industry cooperation and coordination. This paper suggests, then, that one of the principal features of the emerging East Asian approach to the information age is the continued reliance on a corporatist approach. Twenty-first century corporatism--which also exists in European countries such as Germany--can be noted in East Asian nations' practices of collaboration rather than competitive rivalry in the economic, state and labor sectors. In ways that cannot be explained or justified by classical liberal economics, this collaboration now structurally coexists alongside economic liberalization and partial privatization of the information industries in the East Asian region. For this reason, the paper terms the emerging model of the Information Society as the East Asian Liberalized Corporate Model.

Evidence for the model may be noted in the implementation of pro-competition regulatory reforms in the telecommunications industry (OECD, 2000), which has led to universal availability of infrastructure with high penetration rates. In addition, market entry rules for new service providers have been liberalized with few business restrictions, especially in the mobile and wireless communications industry (ibid.). Moreover, most East Asian countries have now passed new laws for cyberspace, especially in the areas of electronic commerce, digital signatures, and data protection (as an example, see South Korean Government, 1999a, 1999b, 1999c).

While the region has been remarkably successful in diffusing new media technologies, especially in the wireless category (see United States Internet Council, 2000; OECD, 2000), it is argued that other policy weaknesses remain in the East Asian approach to the Information Society, stemming largely from the inability to address underlying preconditions and requirements for a knowledge society (see Venturelli, 2001, 1999, 1998a). A few examples of structural weaknesses in the information sector include: a high quality and universal educational system but one that is geared to rote and imitative learning rather than independent, creative, and critical thinking; an underdeveloped regulatory system; and inherent conflicts between the corporatist model of collaboration and the requirements for independent, objective and effective regulation. Although some nations, such as South Korea and Japan may demonstrate some of the highest percentages for Internet usage rates in Asia, the social and economic basis of the information sector is still tied far too much to hardware production and distribution than to the development of advanced information services. Thus we note almost the complete absence of attention in public policy to develop the content industry, which, as argued below (also, see Venturelli, 2001, 2000, 1999) will be the most critical sector in determining wealth creation and social development in the information economy.

***Structural Characteristics of East Asian Policy for the Information Society***

Sometime within the next few years, there is no doubt that the East Asian region will overtake North America and Europe and assume the leadership of the global economy in terms of information technology usage rates. In itself, this will have a profound impact on the social, political, cultural and economic development of East Asian societies. The singularities of the East Asian framework for e-regulation in the Information Society include:

- Continued collaboration between industrial actors and state actors sets upper limits upon the scope of competition in information infrastructure and services
- Persistence of the export model of economic development in information appliances and hardware restrains the pace of domestic consumer exploitation of advanced services
- Tendency to see the Information Society still in industrial terms, bent and shaped to fit the old economy of mass produced electronic hardware commodities
- Missing emphasis on content sector growth, with far too much emphasis on infrastructure and hardware
- Continued reliance on imported technologies, and limited investment in technological innovation or innovation in creative ideas

- Civil society foundations are expanding but are still relatively underdeveloped. There is still scant policy focus on how to promote civil society exploitation of information services without which there will be no explosion in ideas and innovation (see Venturelli, 2001).

What should be stressed is this analysis is that given current trends in policy choices, the East Asian framework for the Information Society will be a distinctive one, with identities all its own. Yet whether these particularities will be beneficial to improving the region's competitive advantage in innovation, creativity, broad social participation in information use, and advanced information service development, remains to be seen. As argued (Venturelli 2001, 1999), investment in hardware, infrastructure, information appliance technologies, and technological diffusion by themselves alone would not resolve the problem of content sector growth, which is the foundation of the knowledge economy.

This examination of the emerging Information Society in the U.S., Europe, and East Asia has attempted to show how each region, given their regulatory legacies, has elected a different mix of good and bad socio-political choices in public policy. Despite the range and diversity of paths to e-regulation suggested in these choices, none adequately addresses the underlying issue of how to promote an innovative society that is open to broad social participation. What is needed at this historical juncture in the international system, it is suggested, is a complete rethinking of the problems and challenges of the new information space in a way that allows its full potential to emerge to the benefit of all sectors of economy and society. The paper concludes, then, with two arguments: first, that the globalization of the information economy and the internationalization of cyberspace have made it imperative that concepts of creativity and innovation be reassessed and repositioned at the center of public policy; and second, that the creative and innovative challenges of the information economy should reform the emphasis on hardware and technological issues by prioritizing the knowledge and creative foundations of society. These arguments suggest the need for a fundamental shift in national and international policy on the appropriate framework for e-regulation in the Information Society.

## Creativity, Innovation and the New Information Space: Shifting the E-Regulation Debate

It argued here (see also Venturelli, 2001, 2000, 1999) that creativity and innovation should be considered as the key to success in the Information Economy, since not merely the sum total of mineral, agricultural, and manufacturing assets, but the ability to create new ideas and new forms of expression comprise an invaluable social resource base. Creative wealth can no longer be regarded in the hereditary or industrial terms of our common understanding, as something fixed, inherited, and mass distributed, but as a measure of the vitality, knowledge, energy, and dynamism in the production of ideas that pervades a given community.

As nations enter the Global Information Society, the greater policy concern should be for forging the right environment (policy, legal, institutional, educational, infrastructure, access, etc.) appropriate to this dynamism as opposed to engaging in a strategic defense of cultural legacy or of the industrial base. The challenge for every nation is not how to prescribe an environment of protection for a received body of cultural or economic capital, but how to construct one of creative explosion and innovation in all areas of the arts and sciences (ibid.). Nations that fail to meet this challenge, it is argued, will simply become passive consumers of ideas emanating from societies that are in fact creatively dynamic and able to commercially exploit the new creative forms.

Several considerations are paramount in the innovation debate. Nation states opposed to the protection of content industries, whether in Europe or elsewhere, are about to discover, if they have not already, that the cultural conflict over media and audiovisual content is not a superficial, high-diplomacy power play between the U.S. and France. It is, instead, about the fate of a set of enterprises that form the core, the so-called 'gold' of the Information Economy (Venturelli, 20001, 2000, 1999). In a feudal, agricultural or a mercantile economy, land, agricultural products, and natural resources such as tea, spices and gold formed the basis of wealth. Gold, in particular, has been the objective currency of wealth across cultures and nations since ancient times. In the industrial age, the basis of wealth shifted to other mineral resources such as oil, and to the creation of capital in plant, equipment, and mass-produced products manufactured from natural raw materials such as iron, oil, and wood. Control over these resources and of the means of transforming them into mass-produced products for distribution to ever wider markets has been the basis of economic power since the industrial revolution. The Information Society is now changing that equation. The source of wealth and power in an information economy is found in a different type of capital: intellectual and creative ideas packaged and distributed in different forms over information networks. On might even say, that wealth-creation in an economy of ideas is derived far less than we imagine from the technological hardware and infrastructure, since eventually most nations, such as China, will make investments in large-scale infrastructure technologies. Rather, this paper suggests, prosperity will be dependent upon the capacity of a nation to continually create content, or new forms of widely distributed expression, for which they will need to invest in creative human capital throughout the economy and not merely in the diffusion of gadgets and hardware.

For instance, every nation will need to have a vibrant audiovisual industry if it is to 'grow' its other multimedia content sectors. In this respect, nations, which attempt effectively to prevent the total erosion of content industries, will have an advantage over those that simply give up the struggle to the inevitable consolidation of international audiovisual producers and distributors. It is no small irony, then, that many countries impervious to the cultural protection argument are now scrambling to find schemes and mechanisms to revive their publishing, film and broadcast sectors, even as they seek ways to encourage the growth and expansion of new content sectors such as software and information services. Mechanisms for protecting and promoting the cultural industries (see Venturelli, 1999, 1998a) include, for example: lottery systems to subsidize film production (UK), taxes on cinema receipts (France), differential postal rates to encourage domestic magazine content (Canada), tax levies on commercial publishers to subsidize small-scale independent publishers (Germany), and structural funds and tax breaks to encourage private investment in content enterprises (Canada, France, Australia, India, among others).

As many nations have yet to discover, the gap in creative productivity does not derive from lower levels of national creative talent or content quality attributes; rather, the paper argues, the gap lies in the power to distribute through advertising, marketing, control of multiple networks, and from horizontal and vertical concentration with other media such as broadcasting, cable, satellite, wireless, and the Internet (Venturelli, 2001, 1998a).

Undoubtedly, the Global Internet will and already is, revolutionizing the manner in which creative forms, including audiovisual products are distributed and consumed. Creative and intellectual property based enterprises and information industries have made this assumption or they would not be actively positioning themselves for the transformation. At the same time, the new information industries are rediscovering the importance of traditional content sectors such as print publishing and film because these enterprises form, it is argued, the creative foundation and feeding line into all the on-line content forms.

In short, a nation without a vibrant creative labor force of artists, writers, designers, scriptwriters, playwrights, painters, musicians, film producers, directors, actors, dancers, choreographers, not to mention engineers, scientists, researchers and intellectuals does not possess the knowledge base to succeed in the Information Economy, and must depend on ideas produced elsewhere.

This changing reality has, in an unexpected way, vindicated the arguments of societies that sought to protect their content enterprises in the name of cultural survival and sovereignty. They were right, though it is suggested here, for the wrong reasons, since it is not the cultural legacy that is at stake, but the capacity to invent and create new creative and innovate forms (see Venturelli, 2000, 1998). Few nations had any notion, even five years ago, that the fate of economy and society would be dependent upon creative resources and the capacity to contribute original forms of expression in the Information Society. From this standpoint, then, all nations will need to regard their content and creative enterprises, including the creative work force, with at least the same value they once ascribed to their metals, mining, minerals, agricultural, heavy manufacturing, electronics, and computing industries. From this standpoint, the approach to e-regulation in the three major economic zones of the world--EU, U.S. and East Asia—is more aligned socio-politically to the regulatory traditions of industrial and post-industrial economies than it is to the basic requirements of a creative economy.

The emergence of ideas as capital, it is argued, necessarily brings creative capital to the center of public policy. The central economic and societal question of the Information Society will soon become: how to stimulate innovation, that is to say, originality in ideas. Through careful and intelligent policy initiatives ranging throughout all social levels, governments will need to provoke a high level of dynamic innovation in the arts, sciences, and imaginative ideas and their integration into an on-line, networked world.

Given the entrenchment of conventional views of communication policy aligned with historically evolved socio-political traditions, it appears unlikely in the medium term that the industrialized nations will re-conceptualize their approaches to the communication and information system. It would require a fundamental shift in approach to knowledge, participation, education, and the social value of the arts and intellectual ideas. Despite grave inadequacies in traditional approaches to thinking about creativity and knowledge in the modern age (see Venturelli 2001, 2000), there may have been little policy incentive historically to reshape the creativity and innovation policy debate and account for its missing dimensions. Yet the information technology revolution has already altered the stakes and made content policy the precondition of how to ensure a creative and innovative society.

What does this challenge involve in terms of public policy? It means an educational system that places emphasis on creative freedom and on incentives for independent thinking; state and private sector investment in research and development of new ideas and technology; and low levels of risk and high levels of reward for creative risk-taking in the workplace and the economy. Most of all, forging an environment of creative dynamism requires regulatory stimulation of creative enterprises, i.e., those enterprises whose products are ideas. It is suggested here that an effective policy framework would, at the very least: (1) broaden access to capital from conventional and unconventional sources; (2) lower taxation on creative risk-taking; (3) remove content obligations and liabilities for all entities that produce and distribute expression; (4) ensure that a constant stream of new ideas and cultural forms trickle into the public domain through 'fair use' access protections; and (5) assure *reasonable*, though not excessive intellectual property rights for innovation in ideas, technology, and science (see Venturelli, 20001, 2000, 1999, 1998a).

Measured by the standards of the role of public policy in promoting favorable conditions for creativity and innovation, none of the three regions examined in this paper can be said to have found the most effective pathway to e-regulation in the Information

Society. It may be fanciful to expect at this time that the international system and leading economies which remain tied to the industrial model of economy and society, will recognize the creative, developmental, and democratic functions of expression and information networks. But it is also inevitable that the Information Society will have to confront the social, cultural, and political effects of profound imbalances and inequalities resulting from ill-conceived policies titled in favor of an industrial view of creative production. Changes in our thinking of what is creativity, innovation, and their basis in the structure of expression, may eventually be forced upon the international system from the high cost some societies will eventually pay for stifling innovation by failing to secure, through appropriate policies, the underlying conditions of a creative economy and a knowledge society.

As the economics of ideas and expression are recognized to play a central and strategic role in everything we do, from politics to banking, from education to consumption, from the organization of the state and the socio-legal system to organization of culture and self-identity, it will become impossible to defend the current design of an information age grounded in industrial economics, and traditional concepts of creativity and knowledge. Whether answering the challenge and closing the gap takes a few years or a century, the historical pressures to revise our approach to these issues is a certainty. Now or in the future, the world's leading economies will one day find themselves on the threshold of an international political settlement to resolve these fundamental principles of a Creative Economy and Information Society. The question is, which nation will transform its domestic policy first and lead the international debate, and which will be surpassed in innovative capacities, forced to spend decades catching up through costly misjudgments.

References

Commission of the European Communities (2000a). Commission gives conditional approval to AOL/Time Warner merger. European Commission press release, Ip/00/1145, October 11, 2000.

__________ (2000b). EC Directive of the European Parliament and of the Council on certain legal aspects of Information Society services, in particular electronic commerce in the Internal Market ("Directive on electronic commerce"), ECO 419/CONSOM 80/CODEC 826, 14263/1/99, 98/0325 (COD).

__________ (1999). Principles and guidelines for the Community's audiovisual policy in the digital age. Communication from the Commission to the Council, the European Parliament, the Economic & Social Committee and the Committee of the Regions, COM(1999)657 final.

_________ (1995a). EC Directive 95/46/EC of the European Parliament and of the Council on the protection of individuals with regard to the processing of personal data and on the free movement of such data, OJL 281/31, 23.11.1995.

_________ (1995b). Proposal for a Council Decision adopting a multi-annual Community programme to stimulate the development of a European multimedia content industry and to encourage the use of multimedia content in the emerging information society, COM (95) 149 final, June 30, 1995.

__________ (1994). 'Europe and the global information society,' Bangemann Task Force Report to the European Council, *Cordis*, Supplement 2, 15 July: 4-31, Brussels: European Commission, DGXIII/D-2.

_________ (1993a). Council Directive on the coordination of certain rules concerning copyright and rights related to copyright applicable to satellite broadcasting and cable retransmission, 93/83/EEC, OJL 248/15, 6.10.93.

__________ (1993b). 'Treaty on European Union (signed in Maastricht on 7 February 1992),' in European Union: Selected Instruments taken from the treaties, book I, vol. I, pp. 11-89. Luxembourg: Office for Official Publications of the European Communities.

__________ (1993c). White Paper on growth, competitiveness, and employment: the challenges and ways forward into the 21st century (the "Delors White Paper"), COM (93) 700 final.

__________ (1989). Council Directive on the coordination of certain provisions laid down by law, regulation or administrative action in member states concerning the pursuit of television broadcasting activities, 89/552/EEC; OJL 298/23, October 17, 1989.

European Parliament (1989). Report on the European Community's film and television industry (the De Vries Report), January 9, 1989, PE 119.192/final.

French Government (2000a). Address by Prime Minister Lionel Jospin at the 20th Summer Forum on Communication, Hourtin, August 26, 1999.

__________ (2000b). "France in the Information Society." Newsletter of the French Government, February 1999, Special Edition.

Godwin, Mike (1998). *Cyber Rights: Defending Free Speech in the Digital Age*. New York: Random House.

Hayek, Friedrich A. von (1944/1994). *The Road to Serfdom*. Chicago: University of Chicago Press.

Hegel, G.W.F. (1821/1952). *The Philosophy of Right*, trans. T. M. Knox. Oxford, UK: Oxford University Press.

Hobbes, Thomas (1651/1991). *Leviathan*, ed. R. Tuck. Cambridge, UK: Cambridge University Press.

Horwitz, Robert B. (1989). *The Irony of Regulatory Reform*. New York: Oxford University Press.

Kant, Immanuel (1784-97/1991). *Political Writings*, ed. H. S. Reiss, trans. H. B. Nisbet. Cambridge, UK: Cambridge University Press.

Killian, Johnny H. and Costello, George A. (eds.), (1996). *The Constitution of the United States of America: Analysis and Interpretation*. Prepared by the Congressional Research Service. Washington, DC: U.S. Government Printing Office.

Kim, K. S. & Roemer, R. (1979). Growth and Structural Transformation. Cambridge, MA: Harvard University Press.

Krol, Ed (1992). *The Whole Internet: User's Guided Catalogue*. Sebastopol, Calif: O'Reilly & Associates.

Lasok, D. and Stone, P. A. (1987), *Conflict of Laws in the European Community*. Abingdon, UK: Professional Books Ltd.

Lessig, Lawrence (1999). *Code and Other laws of Cyberspace*. New York: Basic Books.

Locke, John (1690/1960). *Two Treatises of Government,* introd., P. Laslett. New York: Cambridge University Press.

OECD (2000). Regulatory reform in Korea. Paris: OECD Publications.

Montesquieu, Charles de Secondat (1748/19 89). *The Spirit of the Laws*, trans. A. M. Cohler, eds. B. C. Miller & H. S. Stone. Cambridge, UK: Cambridge University Press.

Nishimizu, M. & Robinson, S. (1984). "Trade Policies and Productivity Change in Semi-industrialized Countries." *Journal of Development Economics*, 16: 177-206.

Rousseau, Jean Jacques (1755/1973). *The Social Contract and Discourses*, trans. G. D. H. Cole. London: Everyman's Library.

Segal, Ben M. (1995). "A Short History of Internet Protocols at CERN." CERN PDP-NS document available at http://wwwinfo.cern.ch/pdp/ns/ben/TCPHIST.html

Smith, Adam (1776/1986). *The Wealth of Nations*, Books I-III, introd. by A. Skinner. London: Penguin.

South Korean Government (1999a). Electronic Commerce Act.

__________ (1999b). Digital Signature Act.

_________ (1999c). Act on the Promotion and Protection of the Information Infrastructure.

Tsao, Y. (1985). "Growth without Productivity: Singapore manufacturing in the 1970s." *Journal of Development Economics*, 18: 25-38.

*Union Pacific Railroad Company v. United States*, 99 U.S. 700 (1978).

U.S. Congress (1998). Digital Millennium Copyright Act, Public Law 105-304, 112 Statute 2860.

________ (1996). Telecommunications Act of 1996, Public Law No. 104-1-4, 110 Stat.56

U.S. Government (1999). "Electronic commerce: Trade policy in a borderless world,"

speech by U.S. Trade Representative Charlene Barshefsky, July 29, 1999.

_________ (1995). Global information infrastructure: Agenda for cooperation. Information Infrastructure Task Force, February. Washington, DC: U.S. Government Printing Office.

________ (1993). The national information infrastructure: agenda for action. Information Infrastructure Task Force, September. Washington, DC: U.S. Government Printing Office.

United States Internet Council (2000). "State of the Internet 2000." Washington, DC: United State Internet Council & ITTA Inc.

Venturelli, Shalini (2001). *From the Information Economy to the Creative Economy*. Forthcoming Monograph from the Brookings Institution & the Center for Arts and Culture, Washington, DC.

________ (2000). "Ownership of Cultural Expression: The Place of Free Speech & Culture in the New Intellectual Property Rights Regime of the European Union." *Telematics & Informatics: An International Journal on Telecommunications & Information Technology*, Special Issue: *The Socio-Cultural Consequences of the European Information Society*, 17(1&2): 9-38.

__________ (1999). "Information Society & Multilateral Agreements: Obstacles for Developing Countries." *Media Development*, Special Issue: Key Issues in Global Communications, 66(2): 22-27.

___________ (1998a). *Liberalizing the European Media: Politics, Regulation & the Public Sphere*. Oxford, UK: Oxford University Press.

__________ (1998b). "Cultural Rights and World Trade Agreements in the Information

Society," *Gazette: The International Journal for Communication Studies*, volume 60(1), pp. 47-76.

____________(1997a). "Information Liberalization in the European Union," *in National Information Infrastructure Initiatives: Vision & Policy Design*, eds. Kahin, B. & Wilson, E, pp. 457-489. Cambridge, MA: MIT Press.

___________ (1997b). "Information Liberalization and the Restructuring of International Relations," in A. Malek and K. Wiegand (eds.), *News Media and Foreign Policy*. Norwood, NJ: Ablex.

Yuhn, Ky-Hyang & Kwon, Jene K. (2000). "Economic Growth and Productivity: A case study of South Korea." *Applied Economic*s, 32: 13-23.